\documentclass[twocolumn,preprintnumbers,amsmath,amssymb]{revtex4}

\usepackage{graphicx}
\usepackage{dcolumn}
\usepackage{bm}
\usepackage{multirow}

\usepackage{color}
\usepackage{amsmath}
\usepackage[colorlinks]{hyperref}
\usepackage[normalem]{ulem}

\begin{document}

\title{Two-dimensional semiconductors in the regime of strong light-matter coupling}

\author{Christian Schneider$^1$}
\author{Mikhail M. Glazov$^2$}
\author{Tobias Korn$^3$}
\author{Sven H\"ofling$^{1,4}$}
\author{Bernhard Urbaszek$^5$}

\affiliation{%
$^1$Technische Physik and Wilhelm Conrad R\"ontgen Research Center for Complex Material Systems, Physikalisches Institut, Universit\"at W\"urzburg, Am Hubland, D-97074 W\"urzburg, Germany\\
$^2$Ioffe Institute, 194021 St.\,Petersburg, Russia\\
$^3$Institut f\"ur Experimentelle und Angewandte Physik, Universit\"at Regensburg, D-93040 Regensburg, Germany\\
$^4$SUPA, School of Physics and Astronomy, University of St. Andrews, St. Andrews, KY 16 9SS, United Kingdom \\
$^5$Universit\'e de Toulouse, INSA-CNRS-UPS, LPCNO, 135 Avenue de Rangueil, 31077 Toulouse, France\\}

\begin{abstract}
The optical properties of transition metal dichalcogenide monolayers are widely dominated by excitons, Coulomb-bound electron-hole pairs. These quasi-particles exhibit giant oscillator strength and give rise to narrow-band, well-pronounced optical transitions, which can be brought into resonance with electromagnetic fields in microcavities and plasmonic nanostructures. Due to the atomic thinness and robustness of the monolayers, their integration in van der Waals heterostructures provides unique opportunities for engineering strong light-matter coupling. We review first results in this emerging field and outline future opportunities and challenges.
\end{abstract}

\maketitle
Transition metal dichalcogenides (TMDCs) are ideally suited as the active material in cavity quantum electrodynamics, as they interact strongly with light at the ultimate monolayer limit. They exhibit pronounced exciton resonances even at room temperature owing to the exceptionally high exciton binding energies of a few 100~meV~\cite{He:2014a,Chernikov:2014a}.\\
\indent The high exciton oscillator strength leads to absorption of up to $20\%$ per monolayer \cite{Li:2014a}, radiative exciton lifetimes on the order of few 100 fs to several ps \cite{Poellmann:2015,Robert:2016a,Moody:2015,Korn:2011a}. In TMDC monolayers (MLs) the dipole selection rules are valley-selective, i.e., distinct valleys in momentum space can be addressed by photons with left- or right-handed helicity \cite{Cao:2012a,Xiao:2012a,Mak:2012a,Sallen:2012a,Zeng:2012a}. In combination with strong spin-orbit splitting this allows studying intertwined spin-valley dynamics of excitons \cite{Plechinger:2016a,Wang:2014b,yang:2015a}. These unique optical properties make monolayer TMDCs, which can readily be embedded in van der Waals heterostructures containing multiple active layers \cite{Geim:2013a,Withers:2016}, ideal systems for investigating excitons and their interactions with other electromagnetic excitations.

This review paper is structured as follows. First, we provide a concise description of the optical properties of excitons in TMDC monolayers. We then present the generic concept of strong light-matter coupling which arises for excitons confined in the TMDC monolayer interacting with photons trapped inside a cavity or plasmons localized in a metallic nanosystem. Strong light-matter coupling gives rise to half-light --- half-matter quasi-particles, which are also known as exciton-polaritons \cite{Weisbuch:1992a,microcavities,Sanvitto:2012a}. This generally results in a substantial modification of the emission properties yielding an oscillatory behaviour between light- and matter excitations in the temporal, and the emergence of the characteristic Rabi-splitting in the spectral domain.  
We review recent experimental advances of strong light-matter coupling in TMDC monolayers and discuss complementary system implementations which were designed to study the formation of exciton-polaritons with atomic monolayers.

\textbf{Excitonic and optical properties of transition metal dichalcogenide monolayers}

\begin{figure*}
\includegraphics[width=0.87\textwidth]{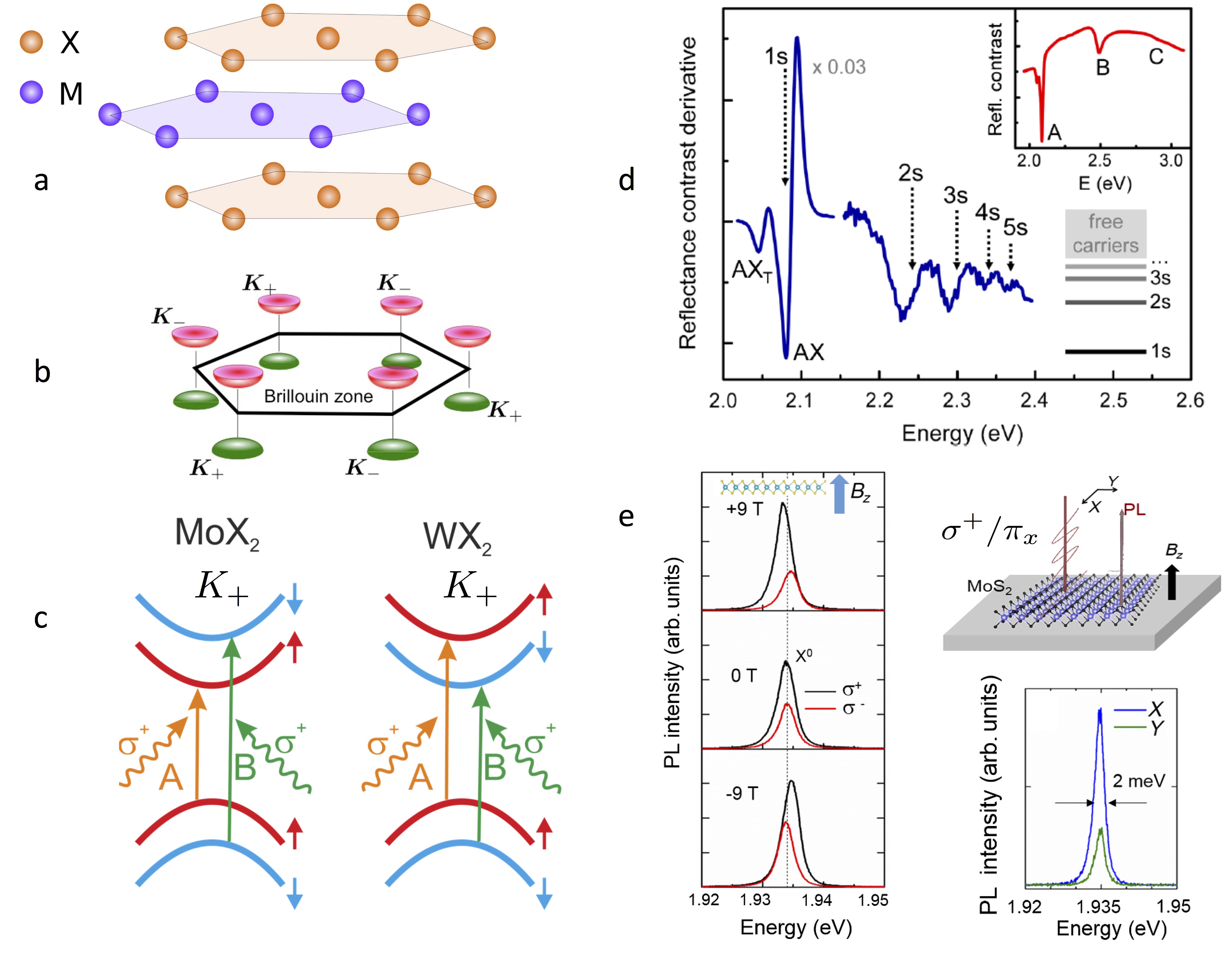}
\caption{\label{fig:fig1} \textbf{Crystal and band structure of semiconducting TMDCs} (a)  Schematic view of TMDC monolayer crystal structure. (b) Band-structure of TMDC monolayer with direct optical bandgap at K-points. (c) Valley specific selection rules for molybdenum-based (MoX$_2$ and tungsten-based (WX$_2$) compounds. (d) Evolution of exciton states for WS$_2$ monolayer on SiO$_2$ \cite{Chernikov:2014a}. (e) Optical valley initilization and valley coherence generation for a MoS$_2$ monolayer encapsulated in hBN \cite{Cadiz:2017a}. }
\end{figure*}

Semiconducting transition metal dichalcogenides are part of the large group of layered materials widely investigated for fundamental research and applications following the discovery of graphene~\cite{Novoselov:2004a}. The remarkably simple mechanical exfoliation techniques give access to rather large-area monolayer samples. While exfoliation of TMDC monolayers was already demonstrated in a seminal work by Novoselov et al.~\cite{Novoselov:2005a} in 2005, the observation of pronounced photoluminescence in MoS$_2$ monolayers, reported by two groups~\cite{Mak:2010a,Splendiani:2010a} in 2010, triggered intense research activities regarding the optical and electronic properties of atomically thin TMDCs.
Monolayers of MoS$_2$ and  related TMDCs consist of a hexagonally coordinated transition metal atom layer sandwiched between top and bottom chalcogen layers, which are also hexagonally coordinated, leading to a trigonal prismatic crystal structure~\cite{Dickinson:1923a,wilson:1969} (see Fig.~\ref{fig:fig1}a) described by the $D_{3h}$ point symmetry group. Correspondingly, the monolayer does not have inversion symmetry. The bulk TMDC crystal is formed by van-der-Waals-mediated stacking the monolayer units. In the 2H stacking sequence, which is the most prevalent polytype, inversion symmetry is recovered for even numbers of layers and eventually in the bulk crystal.

Bulk MoS$_2$ is an indirect-gap semiconductor with a valence band maximum at the $\Gamma$ point, the center of its hexagonal Brillouin zone, and conduction band minima located in between the $\Gamma$ and the $K$ points at the corners of the Brillouin zone. In the monolayer limit, however, the character of the band gap changes to a direct gap at the $K$ points (see Fig.~\ref{fig:fig1}b)~\cite{Li:2007,Mak:2010a,Splendiani:2010a,Kuc:2011a}. 
A similar transition of the band structure from indirect to direct also occurs in the related TMDCs WS$_2$, MoSe$_2$, WSe$_2$, MoTe$_2$ and their alloys. In the TMDC monolayers, the band structure at the K valleys is characterized by a very large, valley-contrasting spin splitting in the valence bands, whose magnitude ranges from about 150~meV (MoS$_2$) to more than 450~meV (WSe$_2$), and a smaller, yet still substantial spin splitting in the conduction band~\cite{Ramasubramaniam:2012a,Dery_FlexPhonons,Kormanyos:2015a}. As the optical transitions between the valence and conduction band are spin-conserving, this splitting gives rise to two, spectrally well-separated interband optical transitions identified as A (transition from the upper valence band) and B (transition from the lower valence band), see Fig.~\ref{fig:fig1}c.

The optical properties of TMDCs are determined by the formation of tightly bound exciton states, which have binding energies on the order of several hundred meVs~\cite{Chernikov:2014a,He:2014a,Ugeda:2014a}, making them stable well beyond room temperature. The large binding energies arise due to a combination of several effects: electrons and holes at the $K$ points of the Brillouin zone have rather large effective masses (ranging from about 0.25~$m_e$ to 0.6~$m_e$ depending on the specific TMDC~\cite{Kormanyos:2015a} where $m_e$ is the free-electron mass) and are strictly confined to the two-dimensional plane of the monolayer. Additionally, their Coulomb interaction is only weakly screened, and this screening typically is anisotropic due to the anisotropic dielectric environment ~\cite{Rytova:1967,1979JETPL..29..658K}. This leads to a strong deviation of excited exciton state energy from a hydrogen-like Rydberg series, illustrated in Fig.~\ref{fig:fig1}d~\cite{Chernikov:2014a,He:2014a}. It is worth noting, that engineering the dielectric environment of the monolayer, e.g., by encapsulating the TMDC between other layered materials, or modifying the substrate, allow for a controlled tuning both of the band gap and the exciton binding energy~\cite{Raja:2017a,Stier:2016b}. The large radiative decay rate of excitons $\Gamma_0\gtrsim 1$~ps$^{-1}$ and, correspondingly, high oscillator strength $f=\Gamma_0/\omega_0 \gtrsim 10^{-3}$, with $\omega_0$ being the exciton resonance frequency, results in efficient light-matter interactions in TMDC monolayers. The exact values of $f$ and $\Gamma_0$ will also depend on the dielectric environment~\cite{Poellmann:2015,Robert:2016a,Moody:2015,Korn:2011a,Jakubczyk:2016a,Ivchenko:2005a}. The short radiative lifetime yields a significant homogeneous spectral broadening of the excitonic transitions~\cite{Selig:2016}. It also leads to a large coupling constant $g$ with photonic modes in microcavity structures, as detailed below \cite{savona95b}.
The high oscillator strength of the excitonic transitions gives rise to a very large absorption for the TMDC monolayer, reaching 20~\% for resonant excitation of the A exciton transition in the tungsten-based TMDCs~\cite{He:2014a,Ye:2014a}. Theoretically, the maximal absorbance of a monolayer $A_{max}$ at resonance is controlled by the ratio of the radiative to the nonradiative, $\gamma$, decay rate of the excitons, $A_{max} = 2\Gamma_0\gamma/(\Gamma_0+\gamma)^2$ and may reach $50\%$ under optimal conditions of $\Gamma_0=\gamma$.  While the emission from typical TMDC samples deposited on SiO$_2$ is strongly inhomogeneously broadened by adsorbates and substrate-induced effects, recent advances in sample fabrication (encapsulation in hexagonal BN) yield linewidths indeed approaching the homogeneous limit, see Fig.~\ref{fig:fig1}e~\cite{Cadiz:2017a,Wierzbowski:2017a,Ajayi:2017a}.

The  transition metal atoms of the TMDCs strongly influence not only the magnitude of the spin splitting, but also the ordering of the spin-split conduction bands (see Fig.~\ref{fig:fig1}c). While for MoX$_2$, the optically bright A-exciton transition connects the upper valence band with the lower conduction band, the band order is opposite in the tungsten-based materials, so that the A-exciton transition addresses the upper conduction band~\cite{Dery:2015a,Kormanyos:2015a}. Thus, for WX$_2$ monolayers, the  exciton state lowest in energy with the electron residing in the lower spin-split conduction band is forbidden in optical transitions for normal light incidence.
The splitting between the optically bright and dark states is given by a combination of the conduction-band spin splitting and electron-hole Coulomb exchange interaction~\cite{Echeverry:2016a}. The lower-energy dark A-exciton state in the tungsten-based materials was indirectly inferred from temperature-dependent PL measurements~\cite{Zhang:2015d,Arora:2015b,Godde:2016a}.  More recently, PL emission from the dark state was directly observed using applied in-plane magnetic fields~\cite{Zhang:2017a,Molas:2017a} and in-plane excitation and detection geometry~\cite{Wang:2017b,Zhou:2017a,ParkKD:2018a}. 
In addition to neutral excitons, charged excitons (trions)~\cite{Mak:2013a} with binding energies of about 25~meV are observable in optical spectroscopy, and the multi-valley band structure allows for different trion species~\cite{Yu:2014b,Jones:2015a,Plechinger:2016a,Courtade:2017a}. Four-particle complexes, biexcitons, i.e., excitonic molecules have also been observed~\cite{You:2015a,Plechinger:2015a,He:2016a}.

The optical selection rules for interband transitions ~\cite{Xiao:2012a} allow for valley-selective excitation at the $K^+$ or $K^-$ valleys using $\sigma^+$ or $\sigma^-$-polarized light, respectively. Thus, near-resonant, circularly polarized excitation generates a valley polarization of excitons, which can be read out directly in helicity-resolved photoluminescence. Even in time-integrated (cw) photoluminescence measurements, large valley polarization degrees are observable for most TMDC monolayers~\cite{Cao:2012a,Xiao:2012a,Mak:2012a,Sallen:2012a,Zeng:2012a,Baranowski:2017a}. These initial observations motivate the  use of the valley pseudospin in potential device applications (valleytronics) \cite{Schaibley:2016a}.  However, the large  cw valley polarization values are, in part, a consequence of the ultrashort exciton radiative lifetime limiting the time window for valley polarization decay. The dominant decay mechanism for excitonic valley polarization is long-range electron-hole exchange interaction~\cite{Glazov:2014a,Yu:2014a}.  Its efficiency scales with the exciton center-of-mass momentum and the resulting decay rate can rival the exciton radiative lifetime, dependant on excitation conditions. In contrast, valley polarization lifetimes are orders of magnitude longer for dark excitons~\cite{Plechinger:2016a,Volmer:2017a}, interlayer excitons in TMDC heterostructures~\cite{Rivera:2016a} and resident carriers in doped TMDC monolayers~\cite{yang:2015a,Dey:2017a}.

\textbf{General framework of strong light-matter coupling}

\begin{figure}[htb]
\includegraphics[width=\linewidth]{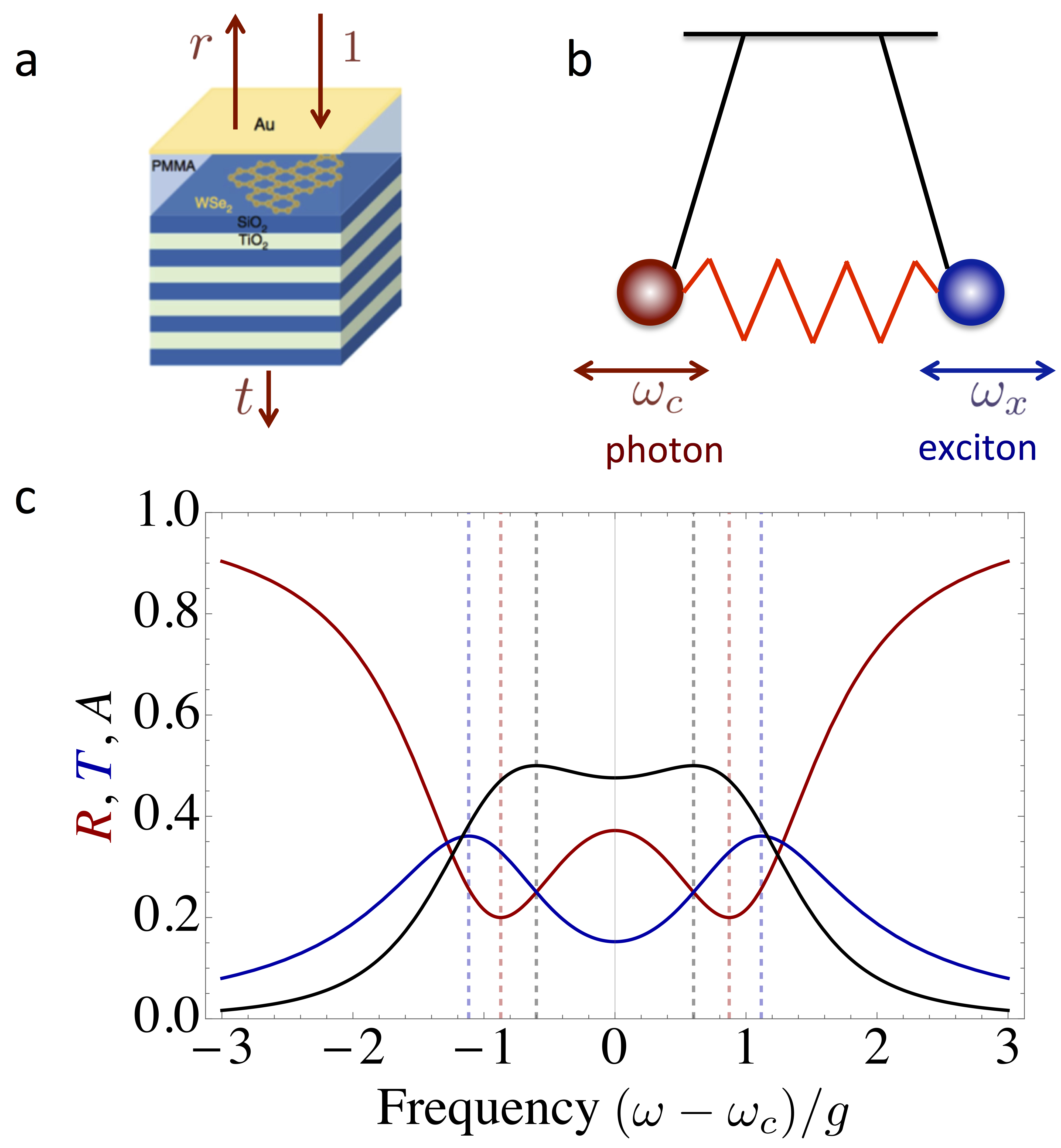}
\caption{\label{fig:fig3mg} \textbf{Basic concept of strong-coupling.} (a) Sketch of the TMDC ML in an optical microcavity fabricated of the distributed Bragg reflector (bottom layers) and metallic layer on top. (b) Illustration of the coupled-oscillators model describing the coherent energy transfer between the cavity photon and exciton. (c) Reflection ($R=|r|^2$, red curve), transmission ($T=|t|^2$, blue curve) and absorption ($A=1-R-T$, black curve) coefficients calculated for the structure in (a) for the strong-coupling parameters $\omega_c=\omega_x$ and $\varkappa=\gamma=0.8g$. }
\end{figure}

An optically active exciton in an isolated TMDC ML emits photons into the free space. In addition to the symmetry-imposed valley selection rules described above, the photon emission process obeys energy and momentum conservation laws, making only excitons with small in-plane wavevectors, $|\bm K|<\omega_{x}/c$, i.e., within the light cone,  subject to the radiative processes. Here, $c$ is the speed of light and $\omega_x$ is the exciton resonance frequency, largely determined by the difference between the free carrier band gap and the exciton binding energy. As emitted light propagates away from the ML carrying away the energy~\cite{Andreani:1991a,Glazov:2014a} the exciton experiences radiative damping. Note that excitons with $|\bm K|>\omega_x/c$ are optically inactive and can contribute to the PL only after relaxation towards the radiative cone.
 The situation becomes qualitatively different if the emitted light cannot leave the vicinity of the ML, e.g., if the ML is placed between the mirrors which form an optical cavity, Fig.~\ref{fig:fig3mg}(a), or if a ML is placed in the vicinity of a metallic or dielectric nanoparticle supporting plasmonic or Mie resonances. In such situations the exciton effectively interacts with a localized mode of electromagnetic radiation (or a plasmon) with the frequency $\omega_c$. Hence, the emitted photon can be reabsorbed by the TMDC ML and reemitted again. This emission-absorption process repeats until either the exciton in the ML vanishes due to the scattering or non-radiative process or the photon leaves the cavity, e.g., as a result of the tunneling through the mirrors. If these decay processes are weak enough the excitation energy is coherently transferred between the exciton and the photon (or plasmon) resulting in the \emph{strong-coupling} regime of the light matter interaction and giving rise to a qualitative change of the energy spectrum in the system: instead of independent exciton and photon states new eigenmodes of the system \emph{exciton-polaritons} are formed~\cite{Weisbuch:1992a,microcavities}.
 
There are several approaches to describe theoretically the strong-coupling effects. It is instructive to consider here the coupled oscillators model where the excitonic contribution to the dielectric polarization in the TMDC ML, $\bm P$,  and the electric field of the cavity mode, $\bm E$, are treated on a semi-classical level and are assumed to obey the oscillator like equations of motion:
\begin{subequations}
\label{oscillators}
\begin{align}
\mathrm i \dot{\bm P} = (\omega_x-\mathrm i \gamma) \bm P + g \bm E,\\
\mathrm i \dot{\bm E} = (\omega_c - \mathrm i \varkappa)\bm E + g\bm P.
\end{align}
\end{subequations}
Here, a dot on top denotes the time derivative, $\gamma$ and $\varkappa$ are the dampings, respectively,  of the exciton and cavity mode unrelated to the light-matter coupling (which determine half-width at half maximum of the resonances), and $g$ is the coupling constant which is determined by the system geometry and exciton oscillator strength. For a planar microcavity it can be roughly estimated as $g \sim \sqrt{\omega_c \Gamma_0}$, where the proportionality constant depends on the cavity geometry and structure of the Bragg mirrors, $\Gamma_0$ is the exciton radiative decay rate into empty space. Here the large exciton oscillator strength resulting in large $\Gamma_0$ allowing to estimate values of $\hbar g \sim 10\ldots 50$~meV depending on the system parameters. These high values for $g$ present one of the intrinsic advantages for studying light-matter coupling in TMDC monolayers as compared to nano-structures with transitions with lower oscillator strength. Equations~\eqref{oscillators} can be formally derived from Maxwell equations for the electromagnetic field in the cavity and the Sch\"{o}dinger equation for the exciton wavefunction in the resonant approximation assuming that $\gamma,\varkappa, g\ll \omega_x,\omega_c$~\cite{Ivchenko:2005a}. It follows from Eqs.~\eqref{oscillators} that for the harmonic time-dependence of the polarization and field $\bm P, \bm E\propto e^{-\mathrm i \omega t}$ the eigenfrequency $\omega$ can be found from the simple quadratic equation:
\begin{equation}
\label{quadratic}
(\omega - \omega_x + \mathrm i \gamma)(\omega - \omega_c + \mathrm i \varkappa) = g^2,
\end{equation}
which indeed describes eigenfrequencies of two damped oscillators coupled with the constant $g$. Equation~\eqref{quadratic} can be also derived from the transfer matrix method which describes propagation of electromagnetic waves in the planar structure or by the procedure of the excitonic and electromagnetic field quantization~\cite{savona95b,Ivchenko:96,Ivchenko:2005a,milburn}.

The general solution of Eq.~\eqref{quadratic} is found in many references, e.g.,~\cite{Ivchenko:2005a,Schneider_2016,microcavities}, here we consider the simplest but already instructive case at resonance $\omega_x=\omega_c\equiv \omega_0$, which already allows one to identify the strong and weak coupling regimes. In this case the solutions of Eq.~\eqref{quadratic} read
\begin{equation}
\label{omega}
\omega_\pm = \omega_0 -\mathrm i \frac{\gamma+\varkappa}{2} \pm \frac{\Omega_R}{2}, \quad \Omega_R = \sqrt{4g^2-(\gamma-\varkappa)^2}.
\end{equation} 
Here $\Omega_R$ is the Rabi frequency related to the so-called vacuum-Rabi splitting, $\hbar\Omega_R$, of polariton modes in quantum electrodynamics. In the \emph{strong-coupling} regime the Rabi frequency is real, i.e., 
\begin{equation}
\label{strong}
\mbox{strong coupling}:~g > |\gamma-\varkappa|/2,
\end{equation} 
In contrast, for $g \leqslant |\gamma-\varkappa|/2$ the light-matter interaction is in the \emph{weak-coupling} regime. The strong coupling means that the real parts of the eigenfrequencies are split by $\Omega_R$, while their imaginary parts responsible for the damping are equalized. In the weak coupling, by contrast, the light-matter coupling affects the damping rates giving rise to the Purcell-like enhancement of suppression of the exciton radiative decay~\cite{purcell,khitrova:purcell}. Hence, in the strong coupling regime an \emph{anticrossing} between the photon and exciton modes should be observed, while in the weak coupling the photon and exciton modes in optical spectra cross each other at the variation of the cavity resonance $\omega_c$ (e.g., via the incidence angle) or the exciton resonance $\omega_x$ (e.g., via the sample temperature).

Equation~\eqref{strong} provides a formal criterion of the strong-coupling regime. In realistic systems, however, the damping of polariton modes $(\gamma+\varkappa)/2$ can be comparable to $\Omega_R$, making identification of the Rabi splitting difficult. 
Moreover, the splitting of peaks in different experiments has different amplitudes \cite{savona95b}: In Fig.~\ref{fig:fig3mg}c we compare the cavity reflection coefficient $R$, transmission coefficient, $T$, and absorbance  $A = 1- R- T$  for $\gamma, \varkappa \lesssim \Omega_R$ where these quantities are found within the input-output formalism~\cite{milburn}. Therefore different experiments, also including PL, on the same sample will give different splittings due to strong coupling that are not directly the Rabi splitting, but are related to it, as detailed in \cite{savona95b}.
 
The model discussion above disregards the nonlinear effects related with the exciton-exciton interactions and the oscillator strength saturation. Since excitons are tightly bound in TMDC MLs these effects are somewhat weaker compared with conventional semiconductor quantum wells, particularly, the exciton oscillator strength saturation is controlled by the parameter $n_{exc} a_B^2$, where $n_{exc}$ is the exciton density and $a_B$ is the Bohr radius.

\textbf{Strong coupling in nanostructures with semiconducting 2D active layers}

\textit{Strong coupling of monolayers in all-dielectric microcavities}

\begin{figure*}
\includegraphics[width=0.87\textwidth]{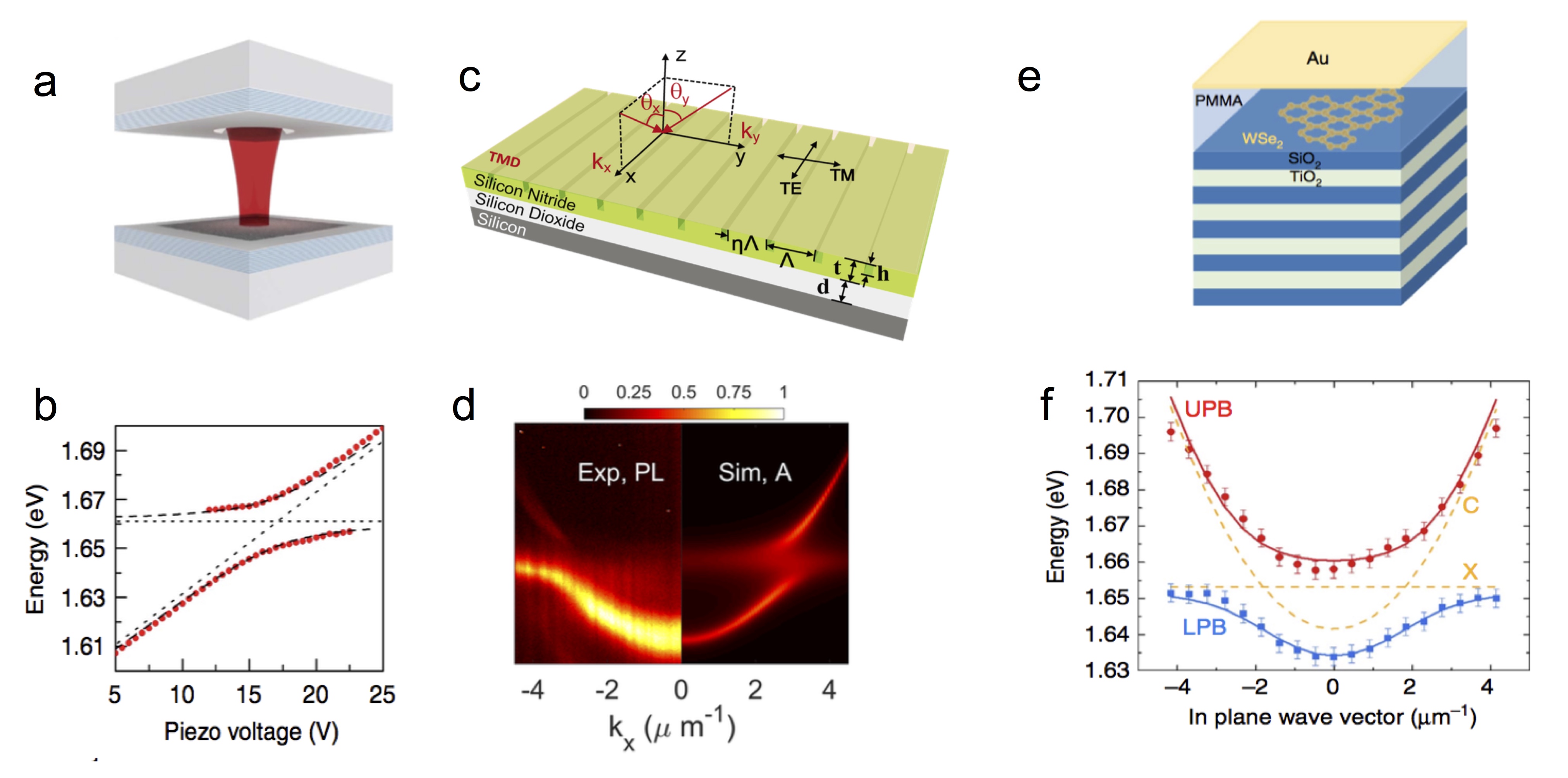}
\caption{\label{fig:fig4} \textbf{Strong coupling in experiments} (a,b) Open cavity based on two separated DBR mirrors \cite{Dufferwiel:2015a}. (c,d) Strong light-matter coupling conditions in lithographically defined second order grating structures \cite{zhang2017room}. (e,f) Hybrid microcavity with DBR at the bottom and thin metal layer on top~\cite{Lundt:2016a} }
\end{figure*}

As described above, the combination of strong exciton binding energies, large oscillator strength and the possibility to strongly reduce structural disorder naturally puts sheets of transition metal dichalcogenides in the focus of polaritonic research. In most III-V, and specifically GaAs-based implementations of polaritonic devices, the cavity design of choice is a high-quality-factor Fabry-Perot resonator based on highly reflecting distributed Bragg reflectors (DBRs), which sandwich the active layer. While, in principle, the transfer of a single, or multiple TMDC layers on-top of a DBR mirror is straight forward, optimal methods to sandwich layers in high-Q DBR-resonators are currently still being developed. This task is closely related to designing and integrating high-quality van der Waals-heterostructures (such as monolayers encapsulated by hBN layers), which reduce inhomogeneous and non-radiative broadening effects dramatically, in more complex devices. Nevertheless, in a first experimental effort, signatures of the strong-coupling regime have been found in a device featuring a single flake of MoS$_2$, synthesized via chemical vapour deposition, that was embedded in a dielectric DBR cavity \cite{Liu:2015b}. There, the authors studied both the reflection spectra as well as the photoluminescence as a function of the in-plane momentum at room temperature. While in this initial experiment, the anticrossing of the normal modes could not be fully mapped out, various groups later on implemented new generations of devices to scrutinize the coupling conditions between confined light-fields and monolayer excitons: A clear cut proof of strong-coupling conditions at cryogenic temperatures has been reported by Dufferwiel et al., for the case of single and double layers of MoSe$_2$, which were embedded in a so-called open cavity based on two separated DBR mirrors~\cite{Dufferwiel:2015a}, see Fig.~\ref{fig:fig4}a,b. In this work, the authors established the formation of exciton-polaritons by fully mapping out the anticrossing of the two resonances in a cavity detuning experiment. A similar implementation, based on an open fiber cavity was reported more recently in \cite{Sidler:2017a}, where strong-coupling conditions were manifested in charge-tunable studies both at the characteristic exciton as well as the trion resonance energies, and the results were interpreted in the framework of coupling to attractive and repulsive polaron resonances. For a monolayer of WS$_2$, the formation of exciton-polaritons was more recently convincingly demonstrated in a fully monolithic cavity in an intermediate temperature regime between 110~K \ldots 230~K \cite{liu2017control}. 
Interestingly, strong light matter coupling conditions in lithographically defined grating structures with a single WS$_2$ monolayer have also been established at room temperature. This approach completely bypasses the difficulties related to capping the atomic monolayer for the integration into microcavities \cite{zhang2017room}, see Fig.~\ref{fig:fig4}e,f.

\textit{Strong coupling of monolayers in metal-based microcavities}

In order to clearly manifest strong-coupling conditions at room temperature with single monolayers, one strategy  involves to replace either one, or both of the DBR mirrors by thin metal layers. In appropriate designs comprising a metal-capped DBR layer, this approach can give rise to so-called Tamm plasmon states, which have significantly reduced mode volumes and thus can be expected to yield increased Rabi splitting \cite{kaliteevski2007tamm,sasin2008tamm}. Strong-coupling conditions with a single monolayer of WSe$_2$ as well as MoS$_2$ in such structures have been reported based on angle-resolved studies. Interestingly, in both efforts, the authors succeeded to map out the full dispersion relation of both the upper and lower polariton branch, as well as the characteristic anticrossing of the normal mode of the system \cite{Lundt:2016a, hu2017strong}, see Fig.~\ref{fig:fig4}c,d. In order to further reduce the effective cavity length and thus increase the coupling strength, approaches involving purely metal based Fabry-Perot-cavities were reported in \cite{Flatten:2016a,wang2016coherent}. While these approaches intrinsically suffer from rather low cavity quality factors (typically $<100$), the observed Rabi-splittings were very large, on the scale of 100~meV. 

\textit{Strong coupling with plasmonic structures}

One route towards further enhancement of the light matter coupling strength is based on plasmonic resonant structures \cite{Cuadra2018}. They also allow to develop truly nanophotonic approaches to confine the light field below the optical diffraction limit. Such devices have been shown to be well compatible with the standard exfoliation and transfer technologies commonly applied in TMDC research. Among the various kinds of available structures, two species of devices have been mostly investigated thus far: The first involves a periodic arrangement (lattices) of metallic nanostructures. Such structures can, e.g. consist of a planar metal layer with holes, or an array of metal disks supporting localized surface plasmon resonances. Here, effects of light-matter coupling have been studied \cite{wang2016coherent, liu2016strong}, and polaritonic behavior was observed.  Nevertheless, while the observed coupling strengths were significant, a clearly resolved split doublet of normal modes was mostly screened by strong broadening effects associated by optical losses in the metal structures. 

The situation becomes even more delicate for systems comprising a single plasmonic resonator coupled to a monolayer: There, it is no longer possible to study the dispersion relation of quasi-particles via angle-resolved luminescence or reflection spectroscopy due to full mode quantization, and the signal strength in standard reflectivity spectra is low. Therefore, a method of choice to investigate light-matter coupling in such systems is so-called darkfield scattering. Thus far, there are a series of reports investigating strong-coupling conditions in TMDC-nanocavity hybrid systems. This includes a demonstration of a two peaked scattering spectrum from a single silver nanorod and a monolayer of WSe$_2$ \cite{Zheng2017manipulating}, respectively a gold rod and a WS$_2$ layer \cite{wen2017room}.   Both reports base their claim primarily on the observation of an anticrossing mode doublet in darkfield scattering spectra, which were acquired by studying a variety of nanorod length to facilitate tuning of the optical resonance frequency. However, it is important to note, that split-peak spectra acquired in darkfield scattering measurements in closely related structures have been priorly convincingly interpreted in the framework of weak coupling: Here, the observed anticrossing is merely a result of enhanced absorption by the excitonic resonance in the presence of a broad optical resonance \cite{kern2015nanoantenna}. 
These ongoing developments indicate the need for complementary experimental evidence to better establish the conditions for observation of strong coupling in these systems. Possible experiments include micro-PL measurements or studies in the time domain.\\

\textit{Polaritons and Valley Selectivity}\\

A unique feature in TMDC based microcavity systems is the possibility to optically address the valley degree of freedom i.e. optical transition in distinct valleys in momentum space \cite{Schaibley:2016a}.  Valley polarization of excitons in WS$_2$, WSe$_2$ and MoS$_2$ is now routinely observed in high-quality samples even under non-resonant excitation conditions \cite{Sallen:2012a,Mak:2012a,Jones:2013a}. In contrast, similar experiments in MoSe$_2$ monolayers only resulted in very low circular polarization of the exciton PL of the order of 5~\% \cite{Wang:2015a}. The dynamic process of valley polarization and depolarization strongly depends on the carrier redistribution, scattering and emission lifetime, and thus it is reasonable to assume that it can be tailored by coupling the excitonic resonances to microcavity modes.

In order to scrutinize whether the effects of valley polarization become more pronounced in strongly coupled microcavities, a variety of experiments have been designed very recently: In \cite{lundt2017valley, dufferwiel2017valley}, the authors have studied a system composed of a single MoSe$_2$ monolayer in the strong-coupling regime with a microcavity mode. Both works independently confirmed, that strong-coupling conditions can retain the valley polarization of the excitations in MoSe$_2$ at cryogenic temperatures, by an amplification of the scattering dynamics. In addition, it was also demonstrated, that the valley index can be directly addressed in the strong-coupling regime by a resonant laser in a Raman-scattering experiment \cite{lundt2017valley}. The great interest to manipulate and enhance spin- and spin-valley related phenomena in the strong-coupling regime, even up to ambient conditions, is further reflected by a series of papers from different groups which demonstrated the valley tagged exciton-polaritons at ambient conditions \cite{sun2017optical, chen2017valley, lundt2017observation} based on monolayers of MoS$_2$ and WS$_2$. These results confirm the great potential of strongly coupled systems to play a crucial role in future valleytronic architectures, where the valley index of monolayer excitons can be married with ultra-fast propagation and low power switching inherited by the polariton nature. \\

\textit{Hybrid Polaritonics}\

In principle, it is possible to generate hybrid states of various excitonic transitions which are coherently coupled to the same photonic mode. These so-called hybrid polaritons have raised considerable interest recently, as they can provide a pathway to combine the advantages of various material systems in one device \cite{Slootsky:2014a}. One example, for instance, involves the case of hybrid structures with embedded semiconductor quantum wells and atomic monolayers. Here, electric current can be injected into one or multiple semiconductor quantum wells which are embedded in a conventional $p$-$i$-$n$ heterostructure. This QW-light emitting diode (LED) can be integrated in the bottom DBR section or into the microcavity. There are two possible processes of coupling between the semiconductor QW and the excitons in the two-dimensional crystals. If coupling between the two excitations is negligible or resonance conditions cannot be established, the semiconductor LED will simply act as an internal light source to excite the excitons in the two-dimensional crystals. However, if strong-coupling conditions can simultaneously be established in the quantum wells and the monolayer crystal with the same photonic resonance, hybrid polaritons composed of excitons in dielectric quantum wells and monolayers can evolve in the system. Such excitations have been observed in \cite{wurdack2017observation} based on a microcavity with four embedded GaAs quantum wells and a single monolayer of MoSe$_2$.  

Light-matter hybridization in the collective strong-coupling regime between monolayer excitons and III-V excitons is also a viable tool to directly influence interactions in the polariton system. It is widely believed, that polariton condensation is strongly facilitated by exciton-exciton exchange interactions, which can yield a stimulated scattering mechanism into a polariton ground state and thus lead to its macroscopic population. This interaction matrix element is given by $\mathcal M = C E_B a_B^2 \sim e^2 a_B/\epsilon_{\rm eff}$ where $E_B \sim e^2/(\epsilon_{\rm eff} a_B)$ is the exciton binding energy evaluated in the hydrogenic model with the effective screening constant $\epsilon_{\rm eff}$ and $C$ is a constant. $\mathcal M$ scales with the excitonic Bohr radius~\cite{tassone,PhysRevB.96.115409}, which is rather small (on the order of 1\ldots 2 nm) in most TMDC materials. Despite somewhat weak dielectric screening in TMDC MLs,  exciton-exciton interaction turns out to be less efficient as compared with III-V semiconductors~\cite{note:M}. By admixing the properties of strongly interacting excitons in III-V materials and strongly bound valley excitons in TMDCs, it is reasonable to believe that a good compromise can be found to facilitate stimulated Bose condensation at elevated temperatures in optimized devices.

Hybrid polariton states were furthermore identified in structures involving organic as well as two-dimensional materials embedded in a fully metallic open cavity \cite{flatten2017electrically}. Such Frenkel-Wannier polaritons should be extraordinarily stable, and represent one promising candidate to observe Bosonic condensation phenomena at strongly elevated temperatures, similar to recent reports on organic-III-V hybrid excitations. \cite{paschos2017hybrid}.

\textbf{Outlook}

Studies of strong light-matter coupling in two-dimensional semiconductors demonstrate outstanding progress \cite{ACSspecial}. By now the strong coupling has been already demonstrated in a number of systems including TMDC MLs in planar microcavities, hybrid organic-inorganic systems, structures with metallic components. \\
\indent First, from a fundamental point of view studies of various collective phenomena and nonlinear phenomena, including possible Bose-Einstein condensation and superfluidity of polaritons \cite{microcavities,Sanvitto:2012a} in atomically-thin semiconductors are very exciting. Since in these systems a truly two-dimensional limit can be realized for excitons, one may expect realizations of novel and previously unexplored facets of complex collective effects in polaritonic systems. \\
\indent Second, both from fundamental and practical viewpoints studies of chirality effects for excitons in two-dimensional materials interacting with light are very promising in view of recent predictions of substantial natural optical activity in TMDC MLs stacks \cite{Poshakinskiy:2017a}. Furthermore, realizations of combined systems with TMDC MLs embedded in chiral cavities open up possibilities of realizing room-temperature circularly polarized lasing~\cite{Demenev2016}.\\
\indent Finally, various applications of strong light-matter coupling for ultrafast optical switching, photon routing and other optoelectronic devices, and, possibly, even for information processing, will naturally appear in the course of further studies of these promising material systems.


\begin{thebibliography}{100}
\bibitem{note:M} The expression for $\mathcal M$ can be also recast via the reduced mass $\mu$ of the electron-hole pair as $\mathcal M \sim \hbar^2/\mu$. Due to larger effective masses in TMDC, the interactions are weaker here that in III-V semiconductor nanosystems.
\end{thebibliography}

\noindent\textbf{Acknowledgements.---} C.S. thanks the ERC for support within the project Unlimit2D. M.M.G. is grateful to the Russian Science Foundation (Grant No. 17-12-01265). T.K. gratefully acknowledges financial support by the German science foundation (DFG) via grants KO3612/1-1 and KO3612/3-1. S.H. is grateful for support within the EPSRC ”Hybrid Polaritonics” Grant (EP/M025330/1). B.U. thanks ANR 2D-vdW-Spin and ERC Grant No. 306719 for financial support.\\

\end{document}